\documentclass[12pt]{article}
\usepackage[utf8]{inputenc}
\usepackage{empheq}
\usepackage{bm}
\usepackage[normalem]{ulem}
\usepackage{latexsym}
\usepackage{color}
\usepackage{amsmath,amsfonts,amssymb}
\usepackage{graphicx,epsfig}
\usepackage{psfrag}
\usepackage{amsthm}
\usepackage{cite,url}
 \usepackage{makecell}

\pdfoutput=1

\usepackage{amsmath}
\usepackage{amssymb}
\usepackage{amsfonts}
\usepackage{mathrsfs}

\usepackage{mathrsfs}
\usepackage{fullpage}
\usepackage{setspace}
\usepackage{bm}
\usepackage{bbm}

\usepackage[normalem]{ulem}
\usepackage{enumerate}
\usepackage{yfonts}

\usepackage{psfrag}

\usepackage{graphicx}
\usepackage{cancel}
\usepackage{slashed}

\usepackage[font=small,labelfont=bf]{caption}
\usepackage{subcaption}

\newcommand{\be}{\begin{equation}}
\newcommand{\bea}{\begin{eqnarray}}
\newcommand{\ee}{\end{equation}}
\newcommand{\eea}{\end{eqnarray}}

\begin{document}
\numberwithin{equation}{section}
{
\begin{titlepage}
\begin{center}

\hfill \\
\hfill \\
\vskip 0.2in

{\Large Shadow analysis for rotating black holes in the presence of plasma for an expanding universe}\\

\vskip .7in

{\large
Abhishek Chowdhuri${}$\footnote{\url{chowdhuri_abhishek@iitgn.ac.in
}}, Arpan Bhattacharyya${}$\footnote{\url{abhattacharyya@iitgn.ac.in}}}

\vskip 0.3in

{\it Indian Institute of Technology, Gandhinagar, Gujarat-382355, India}\vskip .5mm

\vskip.5mm

\end{center}

\vskip 0.35in

\begin{center} {\bf ABSTRACT } \end{center}
We explore the structure of shadow for a Kerr-de Sitter black hole with a non-magnetized, pressureless plasma surrounding it. Specific plasma distributions are considered to separate the Hamilton-Jacobi equation and find the photon regions. An analytic formula describing the boundary curve of the shadow for such a black hole in an expanding universe for an observer at any finite point outside the horizon is derived. We observe deviations which are further explored by calculating the curvature radius at a particular point and distortion parameter for such structures in the presence and absence of plasma and calculate the diameter of the shadow.
\vfill


\end{titlepage}
}

\newpage
\tableofcontents
\section{Introduction}
Black holes have always been the leading contenders when one searches for places where gravity is more robust. They are inevitable consequences of General Relativity and offer insights into the fundamentals of spacetime themselves. Observing such an object can provide us with much-needed data to ensure we are on the right track. However, astronomical evidence assures us that such objects exist. They come from the series of Gravitational Waves detected by LIGO arising out of two black holes merging\cite{Abbott:2016blz,TheLIGOScientific:2016wfe,Abbott:2016nmj,Abbott:2017vtc}. We had strong evidence for the existence of supermassive black holes at the centre of most of the galaxies\cite{Doeleman:2008qh,Dokuchaev:2018kzk,2018,Guo:2018kis} until recently; the ``Event Horizon Telescope"(EHT) has successfully given us a glimpse of such objects\cite{Akiyama:2019cqa,Akiyama:2019brx,Akiyama:2019sww,Akiyama:2019bqs,Akiyama:2019fyp,Akiyama:2019eap}. However, the pictures captured by EHT are not clear enough if we compare them with the precision offered by LIGO and VIRGO. Improving such pictures' sharpness is the reasonable next step by improving the optical interference and telescope technologies. The images are a direct probe of the photon sphere around the black hole. A black hole shadow is formed due to a strong gravitational lensing effect and is defined as the set of directions in the observer's sky from which no signal from the source reaches the observer. The EHT is designed to make such observations possible and gave a plausible picture of the supermassive object's shadow at the centre of our galaxy. 
\par
 Theoretical work on black hole shadows originally began with \cite{The escape} by introducing the idea of escape cone and provided the formula for an angular radius for a static observer at infinity. In \cite{Bardeen}, the author carried out the analysis of rotating black holes and showed that the shadow appears distorted, unlike nonrotating ones. Inspired by these unique optical properties of light rays around black holes, numerous works have been carried out to understand these structures in general relativity and other modified gravity theories\cite{Grenzebach:2014fha,Grenzebach:2015oea,Amir:2016cen,Abdujabbarov:2016hnw,Dastan:2016vhb,Younsi:2016azx,Wang:2017hjl,Cunha:2018acu,Wang:2018eui,Hennigar:2018hza,Ovgun:2018tua,Haroon:2018ryd,Wang:2018prk,Wei:2019pjf,Kumar:2019ohr,Babar:2020txt,Khan:2020hdq,Shaikh:2019fpu,Bambi:2019tjh,Konoplya:2019sns,Contreras:2019nih,Jusufi,Vagnozzi:2019apd,Zhu:2019ura,Ovgun:2019jdo,Contreras:2019cmf,Konoplya:2019goy,Konoplya:2019fpy,Das:2019sty,Lu:2019zxb,Chang:2019vni,Feng:2019zzn,Kumar:2019pjp,Ma:2019ybz,Allahyari:2019jqz,Khodadi:2020jij,Khodadi:2020gns,Kumar:2020hgm,Chang:2020miq, Banerjee:2019nnj,Papnoi:2014aaa,Virbhadra:2008ws,Atamurotov:2013sca}. In most of these calculations, the key point is to define a static observer in a static spacetime or to define a locally static observer in stationary spacetimes. Black holes are not always eternal, and our universe is also expanding. Prompted by this, black holes, in an expanding universe driven only by a positive cosmological constant, has been studied in \cite{Perlick:2018iye}. The black hole considered there was a Schwarzschild black hole, and their work has been extended to a multicomponent universe using an analytic\cite{Tsupko:2019mfo} and approximation method\cite{Bisnovatyi-Kogan:2018vxl}. Further, shadows for the high red-shifted black holes have been studied in these references \cite{Tsupko:2019pzg, Qi:2019zdk, Vagnozzi:2020quf}. These studies have concentrated on nonrotating black holes, but almost all the black holes in our universe are spinning and described by Kerr black holes. In \cite{Li:2020drn,Roy:2020dyy}, the authors have analysed shadow for the Kerr-de Sitter metric. 
\par
For most applications in general relativity, the influence of the medium through which the light rays pass is neglected. However, instances of Solar corona influencing the travel time of radio signals and their deflection angle close to the Sun suggest the presence of medium can give us nontrivial physics. A non-magnetized and pressureless plasma with a linearized gravity theory is enough to describe this physics. In \cite{Muhleman, Muhleman1,B}, authors have focused on the gravitational lensing part. There are good reasons to ask whether the black holes and other compact objects are also surrounded by plasma or not. Some possible effects of plasma in an astrophysical context are reported in\cite{Rogers:2015dla,Rogers:2016xcc,Rogers:2017ofq}. In \cite{Er:2013efa,Virbhadra:1999nm}, authors have studied these effects in a strong lens system. Thus it is an interesting question to explore observable plasma effects for radio signals near compact objects. Studies regarding the influence of a spherically symmetric and time-independent plasma on light like geodesics in Schwarzschild and Kerr spacetime considering deflections on the equatorial plane have been done. In \cite{66}, the authors showed lensing off the equatorial plane, assuming the small deflection, and in \cite{Bisnovatyi-Kogan:2015dxa}, the authors studied plasma effects on gravitational lensing using different methods. They also extended their analysis in the strong bending regime for plasma effects in multiple imaging properties in Schwarzschild spacetime. Furthermore, this kind of lensing has been studied for black holes for gravity theories beyond general relativity \cite{Rahman:2018fgy,Shaikh:2017zfl,Abdujabbarov:2017pfw,Khan:2020hdq}.
\par
Motivated by these, in this paper, we concentrate on the influence of plasma on the shadow of a Kerr-de Sitter black hole. We consider a rotating black hole embedded in an expanding universe: Kerr-de Sitter metric and consider light rays going through such a plasma medium. In \cite{Perlick:2017fio,Yan:2019etp,Atamurotov:2015nra}, the authors discussed the effects of plasma on the shadow near Kerr black holes. Specific plasma distributions are used there, which made the whole study analytically tractable. We give an analytical formula following \cite{Perlick:2017fio} for the boundary curve of the shadow at any arbitrary point outside the horizon of Kerr-de Sitter black hole. The analysis is different from the ones done in \cite{Bardeen} (subsequently extended for Plebanski-Demianski's class black holes in \cite{Grenzebach:2014fha}), where the observer is placed at infinity. 
The distortions of shadow contour are different when we consider homogeneous and inhomogeneous plasma distributions, and we compare them with the well-known shadow structures for Kerr black hole. We study the points that get distorted due to these plasma effects by computing curvatures at those points and the distortion parameters to make a quantitative comparison. 
\par
We organize the paper as follows: In Section~(\ref{sec1}), we find the equations for a light-like particle in non-magnetized, pressure less plasma by providing a suitable Hamiltonian and give the necessary and sufficient condition to the electron plasma density, which separates the Hamilton-Jacobi equation. In Section~(\ref{sec2}), we have discussed the photon region for the plasma density around the Kerr-de Sitter black hole. Knowledge of this photon sphere is essential to describe the boundary curve for the shadow. An observer observes this at any point in the domain of outer communication.  We show our results for various plasma distributions in Section~(\ref{sec5}),  after explaining the method briefly in Section~(\ref{sec4}). In Section~(\ref{sec6}) we make a quantitative study of the results we found in section~(\ref{sec5}) by computing the various observable quantities for such shadows of Kerr-de Sitter black hole in the presence of plasma.
\par
{\bf Conventions:} Choice of signature for the metric is (-,+,+,+). We use units such that $\hbar $=1, i.e., energies have the same unit as frequency, and momentum vectors are the same as wave vectors. All other quantities like c and G have been set to unity.


 \section {Motion of Light rays around the black hole in the presence of plasma}\label{sec1}
We are considering the Kerr- de Sitter metric. It is a solution of the Einstein's equations with a cosmological constant ($\Lambda$) describing a rotating black hole with spin parameter $a$. It is a particular case of the general Plebanski-Demianski family of metrics which are the most general solution of Petrov Type D spacetimes hence making it a general Plebansky-Demiansky metric with zero NUT charge, acceleration, electric and magnetic charges. \par
The metric in Boyer Lindquist coordinates $x=(t,r,\theta,\phi)$ is written as
\begin{equation} \label{met}
ds^{2}=-\frac{\Delta_{r}}{\Sigma}\left[\frac{dt}{\xi}-a \sin^{2}\theta \frac{d\phi}{\xi}\right]^{2}+\frac{\Sigma}{\Delta_{r}}dr^{2}+\frac{\Sigma}{\Delta_{\theta}}d\theta^{2}+\sin^2\theta \frac{\Delta_{\theta}}{\Sigma}\left[\frac{a\, dt}{\xi}-(r^{2}+a^{2})\frac{d\phi}{\xi}\right]^{2}
\end{equation}
with, 
\begin{equation}
  \xi=1+\frac{a^2 \Lambda}{3},\\
  \ \Delta_{r}=\left(1-\frac{\Lambda r^{2}}{3}\right)(r^{2}+a^{2})-2Mr,\ \Delta_{\theta}=1+\frac{a^{2}\Lambda}{3}\cos^{2}\theta,\ \Sigma=r^{2}+a^{2}\cos^{2}\theta.
\end{equation}
The Hamiltonian for light rays around a black hole surrounded by plasma has the following form. Please note that the plasma is considered to be pressureless and non-magnetized \cite{Perlick:2017fio}.
\be{}
H=\frac{1}{2}(g^{{\mu}{\nu}}p_{\mu}p_{\nu}+\omega_{p}(x)^{2}),
\ee
where $g^{{\mu}{\nu}}$is the contravariant form of the metric and $\omega_{p}(x)$ is the plasma frequency of the electron which is related to electron density $N_{e}$ through the relation
\be{}
\omega_{p}(x)=\frac{4\pi e^{2}}{m_{e}}N_{e}.
\ee{}
For details of the derivation of the plasma Hamiltonian interested readers are referred to \cite{SyngeBook}. If we consider $\lambda$ to be a parameter along the curve then the equations governing the light rays are given by
\begin{eqnarray}
\frac{dx^{\mu}}{d\lambda} &=& \frac{\partial H}{\partial p_{\mu}} ;
\label{eom1} \\
\frac{dp^{\mu}}{d\lambda} &=&- \frac{\partial H}{\partial x_{\mu}}.
\label{eom2}
\end{eqnarray}
The light ray trajectories can be found by using the Hamilton-Jacobi equation for the given space-time geometry that we are interested in. It is given by
\begin{equation}
  H\left(x,\frac{\partial S}{\partial x}\right)=0
\end{equation}
where S is the Jacobi action.
The above equation when written in terms of the metric we are using gives \footnote{Our equations are written in Boyer-Lindquist coordinates. This in contrast to \cite{Li:2020drn} where the authors have written the Kerr-de Sitter metric in Boyer-Lindquist coordinates, but in the subsequent analysis, which has been done following \cite{Grenzebach:2014fha}, they have used a rescale coordinate system where $\xi=1+\frac{a^2 \Lambda}{3},$ has been neglected. But in this paper, we have used the Boyer-Lindquist coordinates throughout and kept the factor of $\xi$ in all the equations.}
\begin{align}
\begin{split}
&  \frac{\xi^{2}}{\Delta_{\theta}\sin^{2}\theta}\left(a \sin^{2}\theta\frac{\partial S}{\partial t}+\frac{\partial S}{\partial \phi}\right)^{2}-\frac{\xi^{2}}{\Delta_{r}}\left((r^{2}+a^{2})\frac{\partial S}{\partial t}+a\frac{\partial S}{\partial \phi}\right)^{2}\\&+\Delta_{r}\left(\frac{\partial S}{\partial r}\right)^{2}+\Delta_{\theta}\left(\frac{\partial S}{\partial \theta}\right)^{2}+\omega_{p}^{2}(x)\Sigma=0.
\end{split}
\end{align}
To proceed further we choose the following ansatz expecting the above equation to be separable
\begin{equation}
S=\omega_{0}t+p_{\phi}\phi+S_{r}(r)+S_{\theta}(\theta)
\end{equation}
where $p_{\phi},\omega_{0} $ are conserved quantities which are angular momentum and energy of the test particle respectively.
To further separate the above equation, following \cite{Perlick:2017fio}, we choose our plasma frequency to be of a specific form 
\begin{equation}\label{plasma}
\omega_{p}^{2}(r,\theta)=\frac{f(r)+g(\theta)}{r^{2}+a^{2}\cos^{2}\theta}
\end{equation} and we get,
\begin{equation}
  \frac{\xi^{2}}{\Delta_{\theta}\sin^{2}\theta}(a\ \omega_{0} \sin^{2}\theta+p_{\phi})^{2}-\frac{\xi^{2}}{\Delta_{r}}((r^{2}+a^{2})\omega_{0}+a\ p_{\phi})^{2}+\Delta_{r} (S^{\prime}_{r})^{2}+\Delta_{\theta}(S^{\prime}_{\theta})^{2}+f(r)+g(\theta)=0.
\end{equation}
We can separate out the $r $ dependent and $\theta $ dependent portions and take one of them on the other side to get
\begin{equation}
  \Delta_{\theta} (S^{\prime}_{\theta})^{2}+\frac{\xi^{2}}{\Delta_{\theta}\sin^{2}\theta}(a \ \omega_{0}\ \sin^{2}\theta+p_{\phi})^{2}+g(\theta)=-\Delta_{r}(S^{\prime}_{r})^{2}+\frac{\xi^{2}}{\Delta_{r}}((r^{2}+a^{2})\omega_{0}+a\, p_{\phi})^{2}-f(r)=K
\end{equation}
where $K$ is the generalized Carter constant.
With $S_{\theta}^{\prime}(\theta)=p_{\theta}$ and $S_{r}^{\prime}(r)=p_{r}$ we have the following equations
\begin{equation}
  \Delta_{r}p_{r}^{2}=-K+\frac{\xi^{2}}{\Delta_{r}}((r^{2}+a^{2})\omega_{0}+a\, p_{\phi})^{2}-f(r),
  \end{equation}
  \begin{equation} \label{pt}
  \Delta_{\theta}p_{\theta}^{2}=K-\frac{\xi^{2}}{\Delta_{\theta}}\left( a\ \omega_{0} \sin\theta+\frac{p_{\phi}}{\sin\theta}\right)^{2}-g(\theta).
\end{equation}
We know that $\dot{x}^{\mu}=\frac{\partial H}{\partial p_{\mu}}$ and if we use the Hamilton's Equations then we obtain the following equations for $x^{\mu}=r$ and $x^{\mu}=\theta,$
\begin{equation} \label{eqr}
  \Sigma^{2}\dot{r}^{2}=-K\Delta_{r}+\xi^{2}((r^{2}+a^{2})\,\omega_{0}+a\, p_{\phi})^{2}-f(r)\Delta_{r}=R(r)
  \end{equation}
\begin{equation} \label{eq3}
  \Sigma^{2}\dot{\theta}^{2}=K\Delta_{\theta}-\xi^{2}\left(a\, \omega_{0} \sin\theta+\frac{p_{\phi}}{\sin\theta}\right)^{2}-g(\theta)\Delta_{\theta}=\Theta(\theta)
 \end{equation}
 and we also obtain the equations for $x^{\mu}=t$ and $x^{\mu}=\phi$
 \begin{equation}
   \Sigma\, \dot{t}=\frac{\xi}{\Delta_{\theta}}(p_{\phi}+a\, \omega_{0}\, \sin^{2}\theta)-\frac{\xi}{\Delta_{r}}(r^{2}+a^{2})((r^{2}+a^{2})\,\omega_{0}+a\,p_{\phi})
 \end{equation}
\begin{equation}
  \Sigma\, \dot{\phi}=\frac{p_{\phi}+a\,\omega_0\,\sin^{2}\theta}{\Delta _{\theta}\sin^{2}\theta}-\frac{\xi}{\Delta_{r}}a((r^{2}+a^{2})\,\omega_{0}+a\,p_{\phi}).
\end{equation}


\section{Photon Region} \label{sec2}
To find the shadow curve, we focus on the region of unstable spherical light rays which can serve as a limit beyond which they will spiral down into the black hole. This then will lead to a black patch as none of the light rays reaches the observer's line of sight. Hence this region will serve as the boundary of the shadow. These regions or sometimes they are called photon spheres lead to certain portions of the observer's sky to be dark or unobservable. Thus any unique property regarding the photon region should be reflected in the shadow itself, which will be ultimately captured by EHT. These light rays satisfy $\dot{r}=\ddot{r}=0 $, which are the usual conditions for finding spherical light rays. From equation (\ref{eqr}) we get 
\begin{equation} \label{eq1}
0=R(r)=-(K+f(r))\Delta_{r}+\xi^{2}((r^{2}+a^{2})\omega_0+a\, p_{\phi})^{2},
\end{equation}
\begin{equation}\label{eq2}
  0=R^{\prime}(r)=-(K+f(r))\Delta_{r}^{\prime}(r)-f^{\prime}(r)\Delta_{r}+4\,r\,\xi^{2}\,\omega_0((r^{2}+a^{2})\omega_{0}+a\, p_{\phi}).
\end{equation}
We can solve (\ref{eq1}) and (\ref{eq2}) to get,
\begin{equation} \label{eq4}
  K=\frac{\Delta_r}{(\Delta_r ')^2}\Big[ \left(8\, \xi ^2\, r^2\, \omega_0^2-f'(r) \Delta_r'\right)
  \pm 4 \, \xi\, r\, \omega_0  \sqrt{  \left(4\, \xi ^2\, r^2\, \omega_0 ^2-f'(r) \Delta_r '\right)}\Big]-f(r),
\end{equation}
\begin{equation}\label{eq5}
 a\, p_{\phi}=\frac{ \Delta_r}{\xi\, \Delta_r'}\Big[\xi \, \omega_0  \left(2\, r -\frac{\left(a^2+r^2\right) \Delta_r '}{\Delta_r}\right)\pm \sqrt{ \left(4\, \xi ^2 \, r^2\, \omega_0 ^2-f'(r) \Delta_r '\right)}\Big].
\end{equation}
The left-hand side of equation (\ref{eq3}) being a square can never be negative which leads us to
\begin{equation}\label{eq6}
  (K-g(\theta))\Delta_{\theta}-\xi^{2}\left( a\, \omega_0\, \sin\theta+\frac{p_{\phi}}{\sin\theta}\right)^{2}\geq 0
\end{equation}
and thus
\begin{equation} \label{eq7}
  (K-g(\theta))\Delta_{\theta}\,a^{2}\sin^{2}\theta\geq \xi^{2}\left(a\,\omega_0\,\sin^{2}\theta+a\,p_{\phi}\right).
\end{equation}
Upon inserting the values of $K$ and $a\,p_{\phi}$ from (\ref{eq4}) and (\ref{eq5}), the inequalities (\ref{eq6}) and (\ref{eq7}) with $(r,\theta) $ coordinates tells where the photon region is and also sets the condition where the spherical light rays will be, considering both the plus and minus signs. \par

At this point both the roots in (\ref{eq4}) and (\ref{eq5})  are permitted. Following \cite{Perlick:2017fio}, we can argue that, if we consider the plasma frequency to be very small compared to the photon energy ($\omega_0$), then we can linearize all these expressions in terms of $f(r), g(\theta)$ and their first derivatives.  Now we consider the negative solutions of (\ref{eq4}) and (\ref{eq5}) then it will contradict the fact that  $p_\theta^2$ is positive. This can be easily shown using the equation (\ref{pt}). Hence, from now on, we will only consider the positive root solutions of  (\ref{eq4}) and (\ref{eq5}) i.e
\begin{align}
\begin{split}\label{eq41}
 & K=\frac{\Delta_r}{(\Delta_r ')^2}\Big[ \left(8\, \xi ^2\, r^2\, \omega_0^2-f'(r) \Delta_r'\right)
  + 4 \, \xi\, r\, \omega_0  \sqrt{  \left(4\, \xi ^2\, r^2\, \omega_0 ^2-f'(r) \Delta_r '\right)}\Big]-f(r),\\&
  a\, p_{\phi}=\frac{ \Delta_r}{\xi\, \Delta_r'}\Big[\xi \, \omega_0  \left(2\, r -\frac{\left(a^2+r^2\right) \Delta_r '}{\Delta_r}\right)+ \sqrt{ \left(4\, \xi ^2 \, r^2\, \omega_0 ^2-f'(r) \Delta_r '\right)}\Big].
\end{split}
\end{align}


\section{Constructing the shadow} \label{sec4}
To construct the shadow, we choose a point where the observer will be situated. Let these be $(r_{0},\theta_{0}) $, where $r_{0} $ and $ \theta_{0}$ are the usual Boyer-Lindquist coordinates with $r_{0}> M+\sqrt{M^{2}-a^{2}} $, i.e., outside the outer horizon. Following closely \cite{Grenzebach:2014fha} and \cite{Perlick:2017fio}, we fix tetrads for the observer as 
\begin{align}
\begin{split}
    & e_{0}=\frac{\xi}{\sqrt{\Sigma \Delta_{r}}}(\Sigma+a^{2}\sin^{2}\theta)\partial_{t}+a\partial_{\phi}),e_{1}=\sqrt{\frac{\Delta_{\theta}}{\Sigma}}\partial_{\theta},\\& e_{2}=\frac{\xi}{ \sqrt{\Sigma\Delta_{\theta}}\sin\theta}(\partial_{\phi}+a \sin^{2}\theta \partial_{t}), e_{3}=-\sqrt{\frac{\Delta_{r}}{\Sigma}}\partial_{r} .
\end{split}
\end{align}
Please note these orthonormal set of tetrads are valid only for observers outside the domain of outer communication. For details we refer the reader to look into \cite{Grenzebach:2014fha} and \cite{Perlick:2017fio}.\par
Suppose $\gamma(\lambda) $ are our light rays, the parameter being $\lambda $ along the light ray such \begin{equation} \label{para}g(\dot \lambda,\dot\lambda)=-\omega_p^2(r,\theta).\end{equation} We fix $(r(\lambda),\theta(\lambda),\phi(\lambda),t(\lambda)))$ as the coordinates for the observer position and we trace the light ray by calculating the tangent vector which is 
\begin{equation}
    \dot{\gamma}=\dot{r}\partial_{r}+\dot{\theta}\partial_{\theta}+\dot{\phi}\partial_{\phi}+\dot{t}\partial_{t}
\end{equation}
where the dot means derivative with respect to $\lambda $. On the other hand the tangent vector can also be written as 
\begin{equation} \label{eq8}
    \dot{\gamma}=-\alpha e_{0}+\beta (\sin \tilde{\theta} \cos \psi e_{1}+\sin\tilde{\theta} \sin\psi e_{2}+\cos\tilde{\theta}e_{3})
\end{equation}
where $ \alpha$ and $\beta$ are positive quantities. The $\tilde{\theta} $ and $\phi $ defined in (\ref{eq8}) are the celestial coordinates for our observer, $\tilde{\theta} $ is the co-latitude with its value being 0 for ingoing light rays and $\pi $ for the outgoing ones and $\phi $ being the azimuthal angle. Given the condition in (\ref{para}), from (\ref{eq8}) we can infer
\begin{equation} \label{alpha}
    \alpha^{2}-\beta^{2}=\omega_{p}^{2}|_{(r_{0},\theta_{0})}.
\end{equation}
 $\alpha $ in (\ref{alpha}) can be determined by 
 \begin{equation}
     \alpha=g(\dot{\gamma},e_{0})=\frac{\xi}{ \sqrt{\Sigma \Delta_{r}}}g(\dot{\gamma},(\Sigma+a^{2}\sin^{2}\theta)\partial_{t}+a \partial_{\phi})=\frac{\xi}{ \sqrt{\Sigma \Delta_{r}}}(a\, p_{\phi}+\omega_{0} (\Sigma+a^{2}\sin^{2}\theta)).
 \end{equation}
 $\beta $ on the other hand can be then obtained from (4.4) and thus
 \begin{equation}
     \beta=\sqrt{\frac{\xi^{2}}{\Sigma \Delta_{r}}(a\, p_{\phi}+\omega_{0}(\Sigma+a^{2}\sin^{2}\theta))^{2}-\omega_{p}^{2}}.
 \end{equation}
 All the expressions being evaluated at the observers position  $ (r_{0},\theta_{0}) $ . Now we know our $e_{0}, e_{1}, e_{2}, e_{3} $. Using these in equations (4.2) and (4.3) we can read off the coefficients of $\partial_{r} $ and $\partial_{\theta} $ which will give us a relation between $\tilde{\theta} $ and $\phi $ with $K$ and $a\,p_{\phi} $, where $K$ and $a\,p_{\phi} $ evaluated for the photon sphere. Comparing coefficients of $\partial_{r} $ we get 
 \begin{equation}
     -\beta \cos\tilde{\theta} \sqrt{\frac{\Delta_{r}}{\Sigma}}=\dot{r}.
 \end{equation}
 On squaring both sides we get 
 \begin{equation} \label{eq9}
     \sin\tilde{\theta}=\sqrt{\frac{(K-g(\theta))\Delta_{r}}{\xi^{2}(a\,p_{\phi}+(r^{2}+a^{2})\,\omega_0)^{2}-\Delta_{r}(f_{r}(r)+g(\theta))}}.
 \end{equation}
 Similarly comparing coefficients of $\partial_{\phi} $ we get
 \begin{equation}\label{eq10}
     -\frac{\alpha\, a\, \xi}{ \sqrt{\Delta_{r}\Sigma}}-\beta \frac{\xi}{\sqrt{\Delta_{\theta}\Sigma}\sin\theta}\,\sin\tilde{\theta} \sin\psi=\dot{\phi}
 \end{equation}
 and then inserting all the expressions we get $\sin \psi $ in terms of $K$ and $a\,p_{\phi} $


 \section{Plotting the shadow}\label{sec5}
 The $\sin\tilde\theta $ and $\sin\psi $ expressions will help us in constructing the shadow. At this point we emphasize that, $K$ and $a\, p_{\phi} $ are evaluated at the photon sphere of radius $r_{p},$ whereas $\sin\tilde\theta $ and $\sin\psi $ are evaluated at the position where the observer is present: ($r_{0}$,$\theta_{0}$). The method that we are using here is similar to the one used in \cite{Grenzebach:2014fha} and \cite{Perlick:2017fio}. Interested readers are referred to these references. Here we briefly state them.
 \par
 On putting the expressions for $K$ and $a\, p_{\phi}$ from (\ref{eq41}) in the expressions for $\sin\tilde\theta $ and $\sin\psi $ obtained from (\ref{eq9}) and (\ref{eq10}), the expressions will trace out the shadow boundary. We take the positive root of $K$ to proceed further. The reason for not taking the minus sign is to match with the expressions for low density plasma limit\footnote{For more details we refer the reader to Section VI of \cite{Perlick:2017fio} to understand the low density plasma case}. The boundary we are going to obtain is paramterized by $r_{p} $. The shadow is a closed curve and the parameter $r_{p} $ runs from $r_{p,min} $ which is its minimum value to $r_{p,max} $, its maximum and then again back to $r_{p,min} $. Taking cue from this, we obtain these $r_{p,max} $ and $r_{p,min} $ by setting $\sin\psi=\pm 1. $ These formulas remain valid for any photon frequency $\omega_0$, black hole spin parameter $a$ and any position for the observer $(r_0,\theta_0)$ from the black hole outer horizon.
 \par
 The shadow is drawn using stereo-graphic projection on to a plane that is tangent to the sphere(celestial sphere) at $\theta=0 $. In this plane we use the following Cartesian coordinates 
 \begin{equation} \label{scoord}
   X(r_{p})=-2 \tan\left(\frac{\tilde{\theta}(r_{p})}{2}\right) \sin\psi(r_{p}), \quad Y(r_{p})=-2 \tan\left(\frac{\tilde{\theta}(r_{p})}{2}\right) \cos\psi(r_{p}).
 \end{equation}
 
 This method used in \cite{Grenzebach:2014fha} and \cite{Perlick:2017fio} is different compared to what has been used in \cite{Bardeen} where an observer at infinity uses impact parameters as its coordinates. This method may be compared to the one \cite{Bardeen} only when the observer is at a bigger radius. We briefly lay out the steps for plotting the shadow. For more details, readers are referred to \cite{Perlick:2017fio}.
 \begin{itemize}
   \item Choose the mass parameter M such that $a^{2}\leq M^2$, where $a$ is our spin parameter.
   \item Choose $\omega_p(r,\theta )$ as mentioned in (\ref{plasma}) such that it satisfies the separability condition. to make the equations integrable.
   \item Fix the observer coordinates at ($r_{0}, \theta_{0}$) where $r_{0}$ is the radial coordinate and $\theta_{0}$ is the angular coordinate.
   \item The formulas will have terms like $\frac{\omega_{p}(r,\theta)^{2}}{\omega_{0}^{2}}$. We choose $\omega_{0} $ in units of $\omega_{p}.$ Please note that $\omega_{0} $ is a constant of motion for the light rays.
   \item The coordinates $\sin\tilde \theta $ and $\sin\psi$ which are celestial coordinates are to be written in terms of Carter constant $K$ and $a\,p_{\phi}$ evaluated at ($r_{0}, \theta_{0}$). Note that $K$ and $a\,p_{\phi}$ are evaluated at $r_{p}$, where $r_{p}$ runs over an interval for which unstable spherical light rays exist. Thus $\sin\tilde\theta $ and $\sin\psi$ are functions of $r_{p}$ and them being evaluated at ($r_{0}, \theta_{0}$) and $r_{p}$ running over a certain interval will give us a curve which is the shadow boundary.
   \item To determine the range of $r_{p}$: set $\sin\psi(r_{p})=1$ which will give us $r_{p,min}$ and $\sin\psi(r_{p})=-1$ which will give us $r_{p,max}$. Then we get $\sin\tilde \theta $ and $\sin\psi$, where $r_{p}\ \epsilon\ [r_{p,min},r_{p,max}].$
   \item To get the lower half of the curve we choose $-\frac{\pi}{2}\leq \psi(r_{p})\leq \frac{\pi}{2}$ and $r_{p}$ runs from $r_{p,min}$ to $r_{p,max}$. The upper half is just the mirror image of the lower part about the axis.

 \end{itemize} 
 To proceed further in plotting the shadow we need to know the plasma distributions which satisfies the separability condition in the Hamilton-Jacobi equation. We take cue from \cite{Perlick:2017fio} and choose plasma distributions which describes a homogeneous as well as a non-homogeneous plasma around the black hole. In this paper we mainly use the following distributions:
 
 {\bf Plasma Distributions:}
 \begin{itemize}
 \item \textbf{Case I :} $f(r)=\omega^{2}r^{2} $ and $g(\theta)=\omega^{2}a^{2}\cos^{2}\theta$ such that 
          $ \omega_{p}^{2}(r,\theta)=\omega^{2}\,\, (\omega=\textrm{constant}).$
   \item  \textbf{Case II:} $f(r)=0 $ and $g(\theta)=\omega^{2}M^{2}(1+2\sin^{2}\theta)$ such that $\omega_{p}^{2}(r,\theta)=\frac{\omega^{2}M^{2}(1+2\sin^{2}\theta)}{r^{2}+a^{2}\cos^{2}\theta}.$
   \item \textbf{Case III:} $f(r)=\omega^{2}(M^{3}r)^{1/2} $ and $g(\theta)=0$ such that $\omega_{p}^{2}(r,\theta)=\frac{\omega^{2}(M^{3}r)^{1/2}}{r^{2}+a^{2}\cos^{2}\theta}.$
 \end{itemize} 
 
 \textbf{Case I} describes a homogeneous plasma distribution whereas \textbf{Case II} and \textbf{Case III} describes an inhomogeneous one. The choice of the third case is motivated from power law density behaviours of dust. Also note that, if the plasma frequency $\omega_p(r,\theta)$ is bounded on the domain of outer communication, i.e $\omega_p(r,\theta) \leq \omega,$ light rays with $\omega_0^2 \geq \omega^2$ can travel through any point through that region \cite{Perlick:2017fio}. \par 
 
{\bf Choice of Parameters:} We choose an observer in the equatorial plane such that $\theta_{0}=\frac{\pi}{2}$ and we set $r_{0}=100\,M$ with $M=1$. Since its an expanding universe, $\Lambda=10^{-4}$ and we have chosen $\frac{\omega^{2}}{\omega^{2}_{0}}=1$. To summarize we construct the boundary curve of the shadow for an observer at $r_{0}=100$ and $\theta_{0}=\frac{\pi}{2}$ with a=0.99 \footnote{This choice of $a$ is motivated by the simulation done in EHT paper\cite{Akiyama:2019fyp}, where a highly spinning black hole has been considered.}, $M=1$ and $\frac{\omega^{2}}{\omega^{2}_{0}}=1$. Off-course, our analysis can be done for other choices of parameters. But to establish the key point, the above-mentioned choice is sufficient.  
 We outline the main results obtained below.
 \par 
 {\bf Results:}
 \begin{enumerate}
 \item For the case of homogeneous plasma where $$f(r)=\omega^{2}r^{2}, g(\theta)=\omega^{2}a^{2}\cos^{2}\theta,$$ we compare the Kerr-de Sitter shadow in the presence of such plasma to an environment where there are no such plasma, i.e., $f(r)=g(\theta)=0$. We see the shadow in the \textit{absence} of plasma appears shrunken to that of the shadow in the \textit{presence} of it. The black plot is the shadow for Kerr-de Sitter without plasma while the blue one is in the presence of homogeneous plasma as shown in Fig.~(\ref{fig:fig1}).
 \begin{center}
 \begin{figure}[ht!]
 \includegraphics[width=1\textwidth]{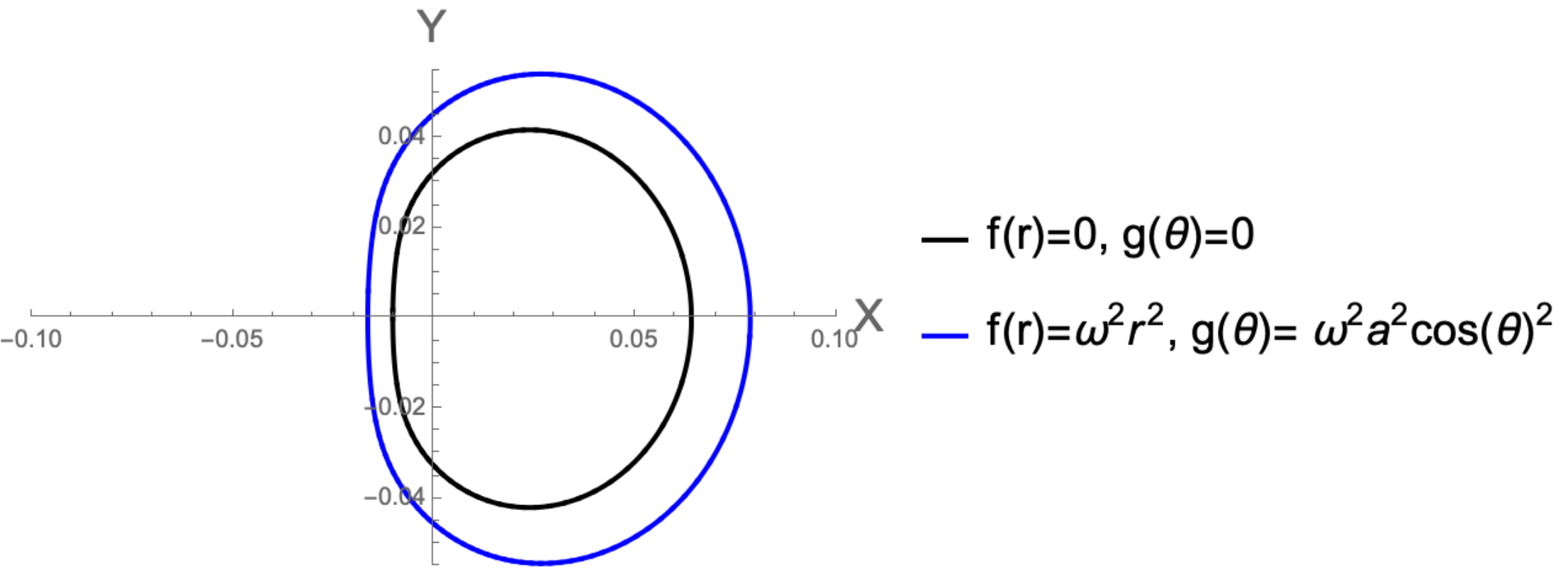} 
\caption{Shadow of Kerr-de Sitter black hole with and without plasma (homogeneous).}
\label{fig:fig1}
\end{figure}
 \end{center}
 \item Secondly, we investigate the shadow of Kerr-de Sitter black hole behaviour in the presence of inhomogeneous plasma environment with that of a without plasma scenario. We find the shadow structure \textit{ almost overlaps (at least the deviation is negligible)} which is in stark contrast to the ones we obtained for the homogeneous ones as in (1). As shown in Fig.~(\ref{fig:fig2}), the black one is again for the Kerr-de Sitter black hole without plasma while the green one represents the shadow in the presence of inhomogeneous media. Here we have used the plasma profile mentioned in {\bf Case III}. Similar conclusion i.e the shadow structures with and without inhomogeneous plasma almost overlap (at least they have negligible difference which we will further quantity in the next section) can be drawn even if we would have used the plasam profile mentioned in {\bf Case II}.
 \begin{center}
  \begin{figure}[h!]
 \includegraphics[width=1\textwidth]{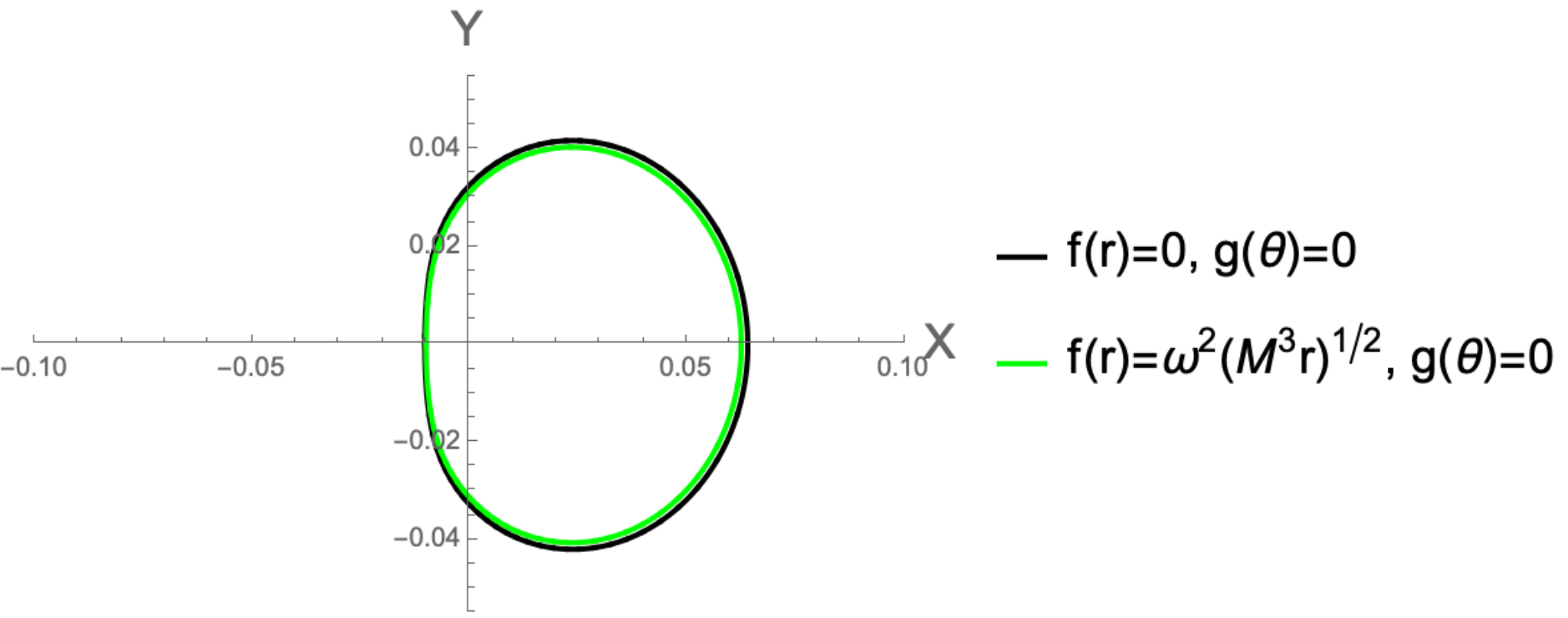} 
 \caption{Shadow of Kerr-de Sitter black hole with and without plasma (inhomogeneous).}
\label{fig:fig2}
\end{figure}
 \end{center}
 \item We compare the behaviour of the Kerr-de Sitter in various mediums- we consider a plasma distribution which is homogeneous and other one is inhomogeneous. For the homogeneous ones $$f(r)=\omega^{2}r^{2}, g(\theta)=\omega^{2}a^{2}\cos^{2}\theta$$ and we consider any one of the inhomogeneous ones mentioned in \textbf{Case II} and \textbf{Case III.} We choose here the \textbf{Case II},  $$f(r)=0, g(\theta)=\omega^{2}M^{2}(1+2\sin^{2}\theta).$$ The key takeaway is that the shadow for \textit{inhomogeneous} plasma environment is smaller than the \textit {homogeneous} one. The red plot is the shadow for the inhomogeneous case while the blue plot is the homogeneous one as shown in Fig.~(\ref{fig:fig3}). Similar conclusion can be drawn using the plasma distribution mentioned in \textbf{Case III}.
 \begin{center}
  \begin{figure}[ht]
 \includegraphics[width=1\textwidth]{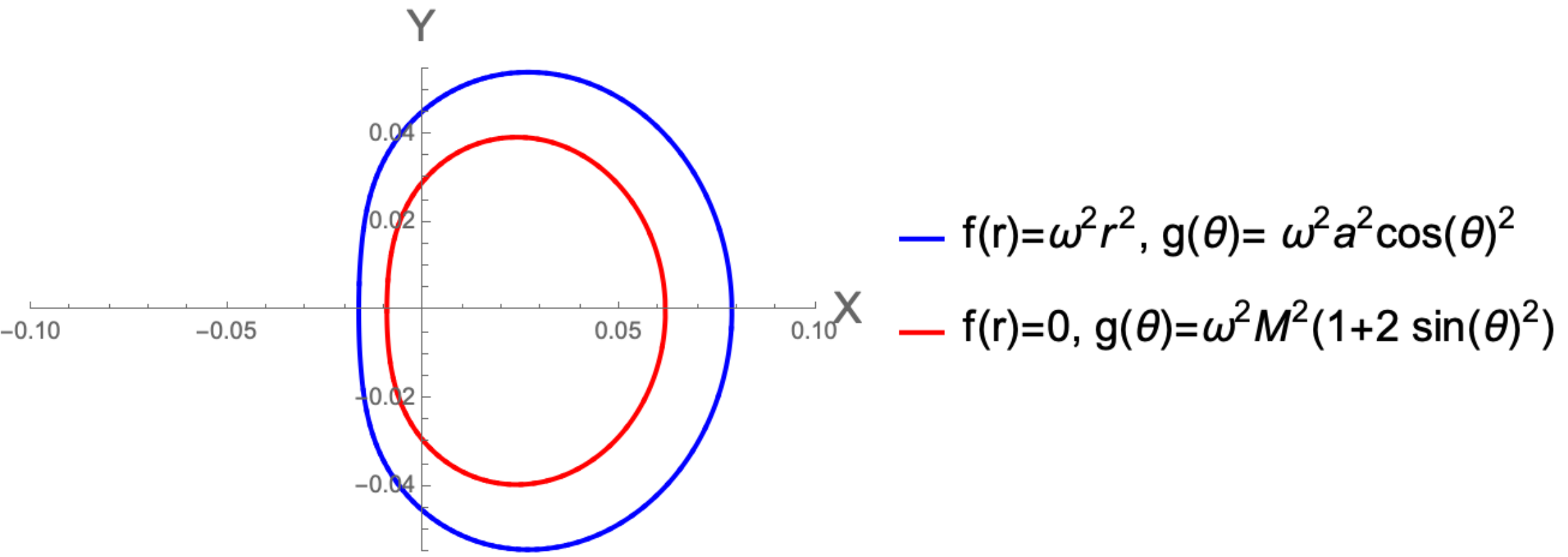} 
  \caption{Shadow of Kerr-de Sitter black hole with  homogeneous and inhomogeneous plasma distributions.}
 \label{fig:fig3}
\end{figure}
 \end{center}
 \item We compare the shadow of the \textit{Kerr black hole without plasma environment} to that of the \textit{Kerr-de Sitter black hole with homogeneous and inhomogeneous plasma environments}. The Fig.~(\ref{fig:fig4}) shows that the size of the shadow of the Kerr-de Sitter black hole with \textit{homogeneous plasma is larger than that of the Kerr metric}.  The black plot is for the Kerr metric shadow without any plasma while the blue one is the Kerr-de Sitter shadow in the presence of homogeneous plasma. Note the parameters chosen are $$f(r)=0, g(r)=0, \Lambda=0$$ for Kerr black hole and $$f(r)=\omega^{2}r^{2}, g(\theta)=\omega^{2}a^{2}\cos^{2}\theta, \Lambda=10^{-4}$$ for Kerr-de Sitter case. 
 \begin{center}
  \begin{figure}[htb!]
 \includegraphics[width=1\textwidth]{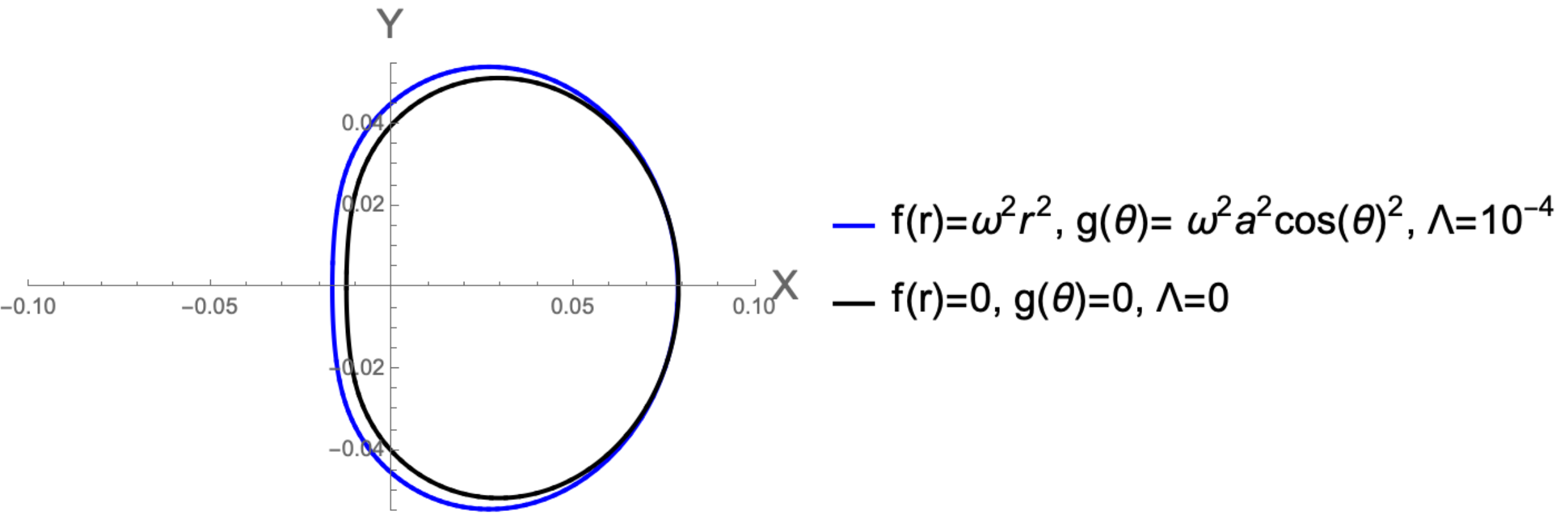} 
 \caption{Shadow of Kerr black hole without plasma and Kerr-de Sitter black hole with  homogeneous plasma distributions.}
 \label{fig:fig4}
\end{figure}
 \end{center}
 However if we consider the inhomogeneous plasma distribution mentioned in {\bf Case III} and $\Lambda=10^{-4}$ we observe that the \textit{shadow becomes smaller from that of Kerr without plasma}. In Fig.~(\ref{fig:fig5}) the black one is still the Kerr metric without any plasma while the red plot is a Kerr-de Sitter metric with an inhomogeneous media around it, the profile of which is stated.
\begin{center}
  \begin{figure}[htb!]
   \includegraphics[width=1\textwidth]{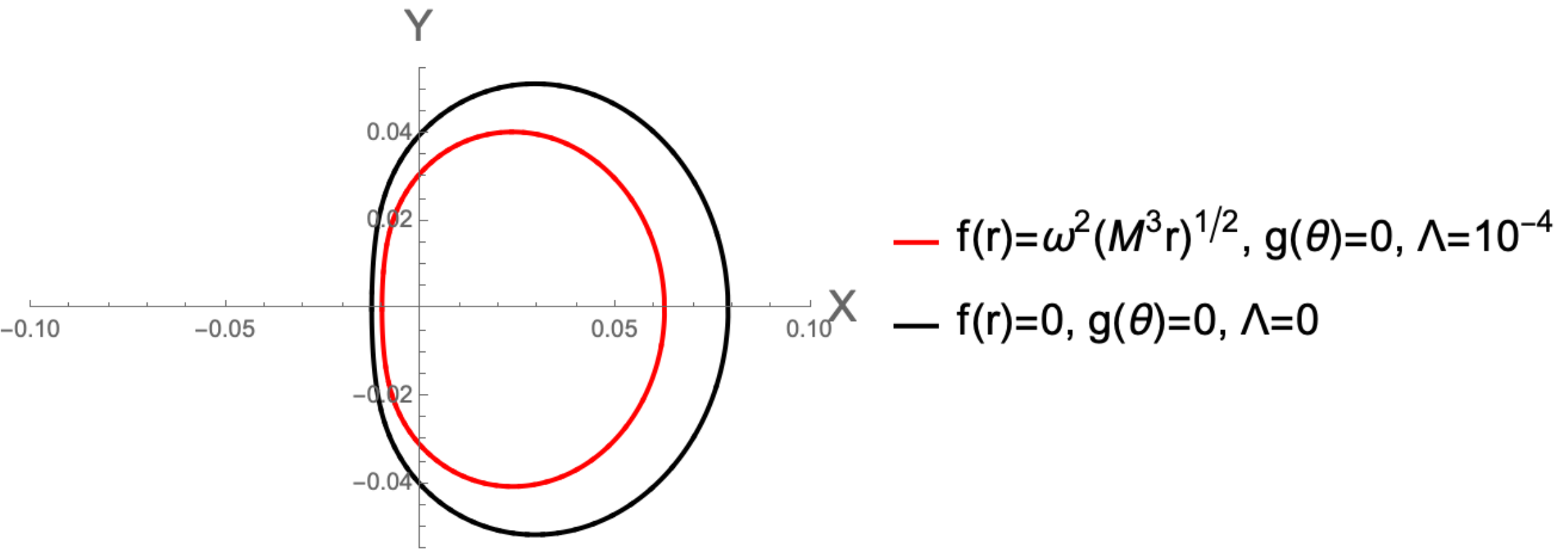} 
\caption{Shadow of Kerr black hole without plasma and Kerr-de Sitter black hole with  inhomogeneous plasma distributions.}
 \label{fig:fig5}
\end{figure}
 \end{center}
 \end{enumerate}
 Similar conclusion can be reached if we use the other inhomogeneous plasma distribution mentioned in \textbf{Case II}.
 
 The plots clearly show that some shadows show distinguishable deviations  in the observer's sky. In contrast, the others are nearly indistinguishable, thus making it difficult to predict the nature of the black hole and its associated environment by solely relying on the shadow image. For the ones where we see an observable deviation, we comment on them by measuring the curvature radius of the boundary curve which we do in the next section.


\section{Observables for the shadow}\label{sec6}
The aim of this section is to study the deviation of the shadows from each other in a more quantitative manner. We compute the observable quantities \cite{Hioki:2009na} namely radius of the shadow and the distortion parameters. 
\par The approach that we take here is in the lines of \cite{Wei:2019pjf, Wei:2018xks, Hioki:2009na} by exploring local curvature radius. 
We know that in two-dimensions, any curve given by a parametric form: $x=x(t) $, $y=y(t) $, t being the parameter along the curve, then the radius curvature at any point $P(x,y)$ is given by 
\begin{equation} \label{coR}
   R=\frac{(x^{\prime2}+y^{\prime2})^{\frac{3}{2}}}{|x^{\prime}y^{\prime\prime}-x^{\prime \prime}y^{\prime}|}.  
\end{equation}
\par
\begin{center}
  \begin{figure}[htb!]
 \includegraphics[width=0.65\textwidth]{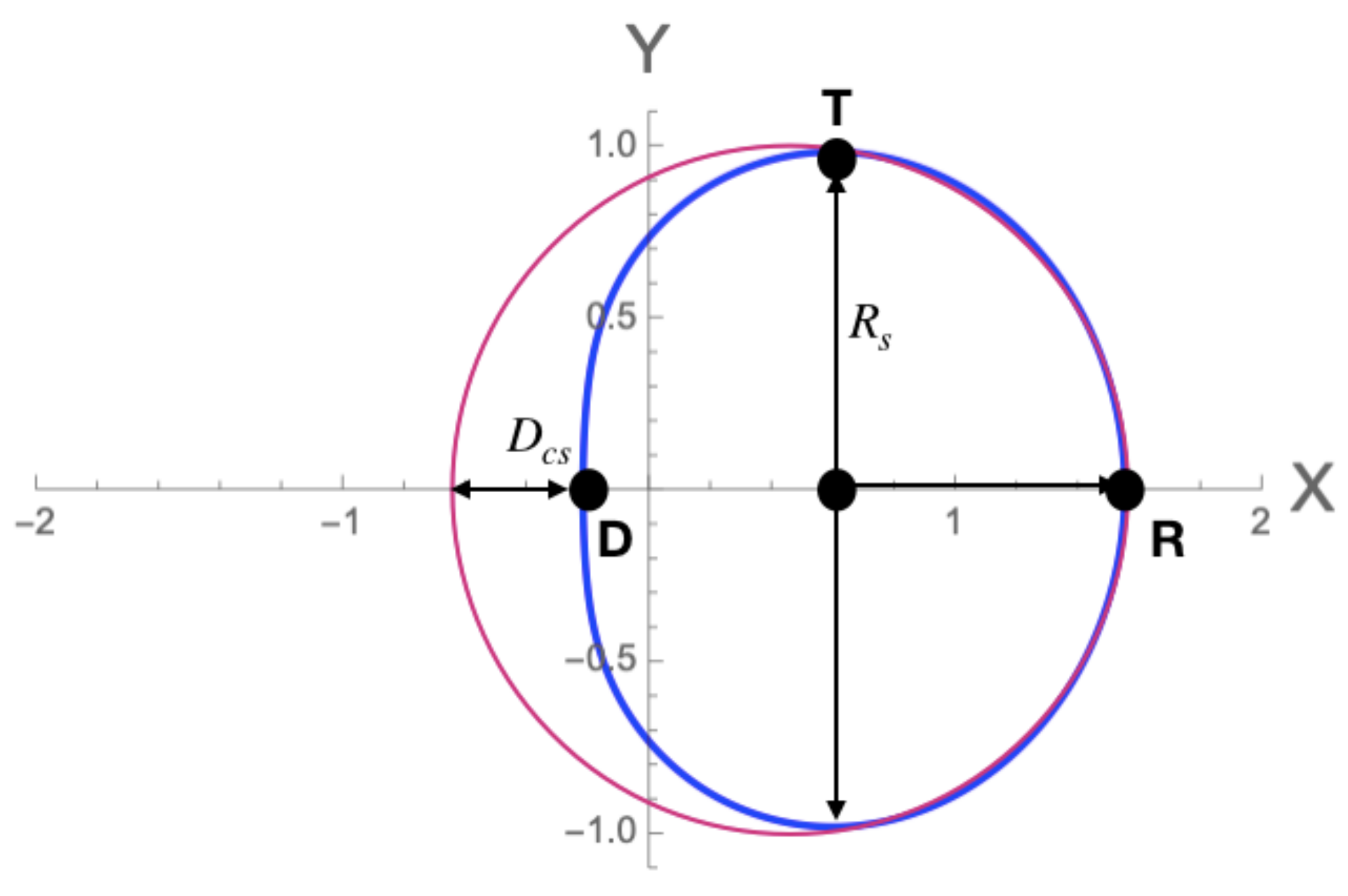} 
 \caption{Figure showing the characteristic points of a typical shadow for a rotating black hole}
 \label{fig:fig6}
\end{figure}
 \end{center}

 
Here we have two coordinates as defined in (\ref{scoord}): $X=X(r_{p})$ and $Y=Y(r_{p})$, $r_{p}$ being the parameter along the boundary curve of the shadow. This point $P(X,Y)$ is on the shadow 
We next choose equatorial plane as before such that $\theta_0=\frac{\pi}{2}.$ 
Given three points we can draw a curve passing through these points. We consider three characteristics points as shown in the Figure~(\ref{fig:fig6}).
\begin{itemize}
    \item Two of the points $D$ and $R$ are defined by $$\sin(\psi)=\mp 1$$ respectively.  $\sin(\psi)$ is defined in (\ref{eq10}).
    \item Other one $T$ (where the coordinate $Y$ takes the maximum value) is defined by $$\partial_{r_{p}}Y|_{\theta_0=\frac{\pi}{2}}=0.$$ 
\end{itemize}
\begin{table}[t!]
\centering
\begin{tabular}{ |c|c|c|c|c|c|c|c|} 
\hline
Shadow & \makecell{Radius of\\ curvature ($R_D$)\\ at point $D$ } & \makecell{Radius of\\ curvature ($R_T$)\\ at point $T$} & \makecell{Radius of\\ curvature ($R_R$)\\ at point $R$}\\
\hline
Kerr BH & 0.149 & 0.046 & 0.053\\
\hline 
Kerr-de Sitter BH & 0.121 & 0.037 &0.043 \\
\hline 
\makecell{Kerr-de Sitter BH \\with plasma: {\bf Case I}} & 0.164 & 0.048 & 0.056\\
\hline 
\makecell{Kerr-de Sitter BH \\with plasma: {\bf Case II}} & 0.088 & 0.035& 0.040\\
\hline 
\makecell{Kerr-de Sitter BH\\ with plasma: {\bf Case III}}& 0.108 & 0.036& 0.041 \\
\hline 
\end{tabular}
\caption{The above table shows the values of curvature radii at a the point $(D,T ,R)$ on the shadow for three different cases (round off to three decimal places). The different cases considered are are Kerr, Kerr-de Sitter without plasma and with plasma. For  Kerr-de Sitter we have set, $\Lambda=10^{-4}$. The plasma frequency chose is $\frac{\omega_{}}{\omega_{0}}=1$. All the quantities are calculated in the unit of $M=1.$}
\label{table:1}
\end{table}

We compute the radius of curvature $R$ defined in (\ref{coR}) at these three points $D,T$ and $R.$ We again set the spin parameter to $a=0.99.$ In the Table~(\ref{table:1}) we list all these radius curvature.\newpage \par
Furthermore, we compute the following two quantities which together with the radius curvature computed before characterises completely the topology of the shadow. 
\begin{itemize}
    \item \textit {The radius of the shadow $R_s:$} It is the radius of an imaginary reference circle relative to the contour of the shadow as shown in the Figure~(\ref{fig:fig6}). For our case it is given by \cite{Hioki:2009na}, 
\begin{equation}
    R_s=\frac{(X_T-X_R)^2+Y_T^2}{2(X_R-X_T)},
\end{equation}
where, $(X_T, Y_T)$ and $(X_R,0)$ are the coordinates of the points $T$ and $R$ respectively. 
\item \textit{The distortion parameter $\delta_s:$} This gives the distortion of the shadow contour with respect to the reference circle \cite{Hioki:2009na}.
\begin{equation}
\delta_s=\frac{D_{cs}}{R_s},\quad D_{cs}=2 R_s-(X_D-X_R),
\end{equation}
where, $(X_R,0)$ are the coordinates of the point $R$ as shown in the Figure~(\ref{fig:fig6}).
\end{itemize}
The values of $R_s$ and $\delta_s$ for various cases are given in the Table~(\ref{table:2}). The values tell us precisely the distortions of the shadow in the presence and absence of the medium. The reason we are computing distortion parameter is to argue our conclusions from a physical standpoint since it is a physical observable for the shadow and indeed EHT data shows a deviation from circularity. \par We compute the deviations using the above definition and list them in Table~(\ref{table:2}). Particularly, one can see the from the last column of the table that the values of the horizontal diameters $(X_R-X_D)$ of the shadow for Kerr-de Sitter Black hole without plasma and with inhomogenous plasma are very close to each other compared to the case with homogenous plasma distribution. Also, the horizontal diameter of the Kerr black hole is smaller than that of the Kerr-de Sitter balck hole with \textit{homogenous} plasma but greater than the Kerr-de Sitter balck hole with \textit{inhomogenous} plasma. Similar conclusions can be reached by looking at the values of the other observables e.g $R_s,\delta_s.$ Hence these support our the claims  made in the  section (\ref{sec5}). 
\begin{table}[t!]
\centering
\begin{tabular}{ |c|c|c|c|c|c|} 
\hline
Shadow & $R_s$ & $\delta_s$&$(X_R-X_D)$ \\
\hline
Kerr BH & 0.051 & 0.224&0.091 \\
\hline 
Kerr-de Sitter BH & 0.042 & 0.224 &0.074 \\
\hline 
\makecell{Kerr-de Sitter BH \\with plasma: {\bf Case I}} & 0.054 & 0.245& 0.095\\
\hline 
\makecell{Kerr-de Sitter BH \\with plasma: {\bf Case II}} & 0.039 & 0.199 &0.071\\
\hline 
\makecell{Kerr-de Sitter BH\\ with plasma: {\bf Case III}}& 0.040 & 0.209 & 0.072\\
\hline 
\end{tabular}
\caption{The above table shows the values of radius $R_s$, the distortion parameters $\delta_s$ and $D_{cs}$ for three different cases. The different cases considered are are Kerr, Kerr-de Sitter without plasma and with plasma. For  Kerr-de Sitter we have set, $\Lambda=10^{-4}$. The plasma frequency chose is $\frac{\omega_{}}{\omega_{0}}=1$. All the quantities are calculated in the unit of $M=1.$}
\label{table:2}
\end{table}
\section{Discussion}
Investigating photon regions in the presence of a nontrivial medium is always of importance. This work is in those lines but takes a more realistic setting where we have an expanding universe with a positive cosmological constant. The plasma has a refractive index, which is frequency-dependent, which makes it a dispersive medium. The presence of plasma influences the trajectories of light rays for the domain of outer communication. Our study is valid for any value of the black hole spin parameter $a $. The medium being dispersive photons of different frequencies and wavelengths have different trajectories. We have used the Hamilton-Jacobi equations and separated the equation by imposing a specific condition on the plasma frequency and obtained the generalized Carter constant. Given the separability condition, we find photon regions satisfying this condition and give an analytical formula for the boundary curve of the shadow. This is done by defining two angular coordinates. \par
The analysis done is valid for any photon frequency, spin parameter, and any observer outside the horizon having an arbitrary inclination and its limiting value being infinity. The presence of plasma introduces changes in trajectories in the light rays and eventually affects the black hole's shadow. Processes like absorption and scattering have not been considered, but the analysis can be extended using the method of \cite{James:2015yla}. The figures in Section~(\ref{sec5}) show the interesting behaviours of the photon ring trajectories in a homogeneous and inhomogeneous media. For Kerr-de Sitter black hole, the shadow gets distorted only when they are homogeneous. \textit{The interesting facts seem to appear that, for inhomogeneous distribution, we don't get much deviation from the one withour plasma. Further, if we compare the shadows in homogeneous and inhomogeneous media, we observe a deviation between them.}  Depending on the nature of the plasma distribution the shadow for Kerr-de Sitter black hole can be smaller or bigger than the corresponding one for Kerr black hole without any plasma. Hence we can conclude that the environment plays a very intricate role in determining the structure of the shadow and further investigation with more realistic plasma distribution should be done. We leave this for future study.  

As for future studies, one can study more complicated scenarios like black holes with accretion disks around them. We have considered an expanding universe here. Studying these similar structures for an observer fixed with the expansion, which is called `co-moving observer' is also a task one can do following \cite{Li:2020drn,Perlick:2018iye}. The presence of jets around them is also a possibility. Exploring these structures by introducing nontrivial terms considering photons interacting with other fields like a scalar field is also interesting. Some works in this line are done recently in \cite{Hu:2020usx,Vrba:2019vqh}, considering QED effects. Doing these analyses for naked singularities is another possibility as they are very different from the usual black hole ones, as shown in \cite{73}.  Last but not the least, it will important to investigate the polarimetric signature of ``photon ring" surrounding the black hole shadow \cite{Himwich:2020msm,Gralla:2019xty,Hadar:2020fda} in presence of plasma. We hope to report some of these studies soon.

\section*{Acknowledgements}

Research of A.C. is supported by the Prime Minister's Research Fellowship (PMRF-192002-1174) of Government of India. A.B is supported by Start Up Research Grant (SRG/2020/001380) by Department of Science \& Technology Science and Engineering Research Board (India). Authors thank Mostafizur Rahman for useful discussions. 

\end{document}